\documentclass[runningheads]{llncs}
\usepackage{graphicx,subfigure,sidecap}
\usepackage{amsmath}
\begin{document}
\title{Generative Artificial Intelligence Reproducibility and Consensus}

%

\author{Edward Kim$^{1}$, Isamu Isozaki$^{1}$, Naomi Sirkin$^{1}$, Michael Robson$^{2}$}
\institute{
$^1$Department of Computer Science, Drexel University, PA\\
$^2$Department of Computer Science, Smith College, Massachusetts, MA\\
{\tt\small ek826@drexel.edu,imi25@drexel.edu,nhs39@drexel.edu,mrobson@smith.edu}
}

\maketitle

\begin{abstract}
We performed a billion locality sensitive hash comparisons between artificially  generated data samples to answer the critical question - can we reproduce the results of generative AI models?  Reproducibility is one of the pillars of scientific research for verifiability, benchmarking, trust, and transparency.  Futhermore, we take this research to the next level by verifying the ``correctness'' of generative AI output in a non-deterministic, trustless, decentralized network.  We generate millions of data samples from a variety of open source diffusion and large language models and describe the procedures and trade-offs between generating more verses less deterministic output.  Additionally, we analyze the outputs to provide empirical evidence of different parameterizations of tolerance and error bounds for verification.  For our results, we show that with a majority vote between three independent verifiers, we can detect image generated perceptual collisions in generated AI with over 99.89\% probability and less than 0.0267\% chance of intra-class collision.  For large language models (LLMs), we are able to gain 100\% consensus using greedy methods or n-way beam searches to generate consensus demonstrated on different LLMs.  In the context of generative AI training, we pinpoint and minimize the major sources of stochasticity and present gossip and synchronization training techniques for verifiability.  Thus, this work provides a practical, solid foundation for AI verification, reproducibility, and consensus for generative AI applications.

\end{abstract}

\section{Introduction}
\label{sec:intro}
Generative Artificial Intelligence (GenAI) represents one of the most impactful and consumer pervasive advancements in artificial intelligence technology in recent years. This form of AI is designed to learn the distribution of the training data, and sample from the learned manifold to create content, e.g. text, images, music, or other complex signals. It a significant shift from traditional AI models that are primarily used to analyze and interpret data typically seen in discriminative supervised tasks such as classification or regression. This raises issues of scientific reproducibility due to the statistical nature of generative AI.  Reproducibility is essential as it enables the independent verification of results, ensuring that findings from a machine learning model or algorithm are valid and reliable.  This is also important for understanding of a model's behavior, its strengths, and weaknesses.  Reproducibility allows for the independent and accurate benchmarking and comparison of ML models through the open sharing of datasets and code; however, given the current trajectory of large ML models, understanding the issues and challenges around reproducibility is increasingly important.

As a case in point, AI is growing exponentially in every aspect - in usage and adoption, as well as in cost to train, model parameterization, data, and compute. For example, ChatGPT reached 1 million users in just 5 days -a feat that took 3.5 years for Netflix, and 2 years for Twitter. OpenAI's GPT-3 required over over \$12M USD in training costs, and the carbon footprint of training this model was equivalent to the output of 126 danish homes for an entire year \cite{anthony2020carbontracker}.  This does not even cover the cost of obtaining and labeling massive amounts of data.  The cost, size, and compute necessary for GPT-4 is an order of magnitude larger, and intractable to train or deploy by anyone except for a handful of the largest industry players.   However, advancements in model quantization, hardware, and cloud computing have made it possible to train and deploy increasingly complex models on consumer grade hardware.  Thus, a significant effort is underway in the field to democratize artificial intelligence.

One of the main goals of the democratization of AI is to expand the benefits of AI beyond a small group of elite researchers and companies, and to ensure that everyone has access to the tools and resources they need to take advantage of AI. While open source software is a core component, there is also the need for ``open source'' hardware.  This concept is not new and has been explored in decentralized cloud networks.  For example, BOINC \cite{anderson2004boinc}, which stands for Berkeley Open Infrastructure for Network Computing, is an open-source software platform for distributed computing. It allows volunteers to donate unused processing power from their personal computers to scientific research projects.  The system is designed to manage and utilize the computing resources of thousands of volunteers across the globe, effectively creating a massive, distributed supercomputer. This allows researchers to conduct large-scale computations without the need for a dedicated, or centralized infrastructure.  Some well-known projects include SETI@home \cite{anderson2002seti}, which searches for extraterrestrial intelligence, and Folding@home \cite{beberg2009folding}, which simulates protein folding for disease research.


However, a decentralized cloud infrastructure must utilize trustless computing principles that do not normally apply when dealing with a known entity. In a decentralized ecosystem, you do not know who you are interacting with, and the dangers of malicious or adversarial actors increases by orders of magnitude. To address these issues, the contributions of this work are the following.  We investigate and present techniques to scale up the computational capabilities of a decentralized network. We present algorithms that can execute ML tasks and monitor other nodes for fraudulent output and, in our results, provide empirical evidence of different parameterizations of tolerance and error bounds for verification.  In the context of generative AI training, we pinpoint and minimize the major sources of stochasticity and present gossip and synchronization training techniques for verifiability.  In essence, this work demonstrates that we can reproduce and verify generative AI work in both image and language generation with minimal overhead and extremely high precision in a decentralized, trustless machine learning network.

\section{Background}
\subsection{Background in Generative AI}
Algorithmically, deep learning currently dominates nearly all applications of artificial intelligence and machine learning, and has shown incredible success in the past several years. These improvements can be attributed to multiple factors, where two major contributors were access to large amounts of data, and large amounts of compute power.  Today, training and running these models requires an enormous amount of compute, usually accelerated in cloud infrastructure using high-powered GPU hardware.

Since the introduction of Generative Adversarial Networks (GANs) \cite{goodfellow2020generative}, generative AI has made remarkable strides. Today, it's used in a wide range of applications, from creating realistic imagery and generating art to synthesizing high-quality speech and writing.  Stable Diffusion \cite{Rombach_2022_CVPR}, is a deep learning, text-to-image model primarily used to generate detailed images conditioned on text descriptions. It uses a latent diffusion model, which involves training the model to remove successive applications of Gaussian noise on training images, functioning as a sequence of denoising autoencoders. The model consists of three parts: a variational autoencoder (VAE) \cite{kingma2019introduction}, a U-Net \cite{ronneberger2015u}, and text encoder \cite{li2017learning}. The VAE compresses an image from pixel space to a smaller dimensional latent space, capturing a more semantic meaning of the image. Gaussian noise is then iteratively applied to this compressed latent representation during forward diffusion. The U-Net block, composed of a ResNet \cite{he2016deep} backbone, denoises the output from forward diffusion to obtain a latent representation. The VAE decoder then generates the final image by converting this representation back into pixel space. This process can be conditioned on a string of text, an image, or another modality.  Large language models (LLMs) are another class of generative AI that have been trained on large textual datasets. These models have shifted the focus of natural language processing research away from training specialized supervised models for specific tasks. Despite being trained on simple tasks such as predicting the next word in a sentence, LLMs with sufficient training and parameter counts capture much of the syntax and semantics of human language and demonstrate considerable general knowledge within their training corpus. 

\subsection{Background in Decentralized Verification and Consensus}
The field of machine learning verification is in its infancy.  In the traditional definition, verification involves making a compelling argument that the system will not misbehave under a broad range of circumstances \cite{xiang2018verification}.   This involves the need to consider unusual inputs crafted by an adversary, not just naturally occurring inputs as testing alone is insufficient to provide security guarantees \cite{goodfellow2017challenge}.

However, we are dealing with a nuanced type of verification of correctness.  In the context of a decentralized network, a consensus model is a mechanism that ensures all participants in a distributed network agree on the content of a shared database. It is the protocol by which the nodes in the network agree on a single version of the truth, despite the presence of faulty nodes or those with malicious intent. The consensus model is the core idea to minimize the need for trust in a blockchain system, where consensus ensures that every transaction is validated according to a set of agreed-upon rules.   Unlike traditional centralized systems where a single authority validates transactions, in a blockchain, multiple nodes participate in the validation process \cite{baliga2017understanding}. This decentralization enhances security and transparency but requires the following properties to be effective \cite{ferdous2020blockchain}. (1) Agreement - all honest nodes must agree on the same value. (2) Validity - if all honest nodes propose the same value, they must decide on that value. (3) Termination - every honest node must eventually reach a decision. (4) Integrity - a node decides on a value at most once. In other words, once a node has made a decision, it cannot change it. And (5) Fault Tolerance - the consensus model should be able to function correctly even if some nodes fail or act maliciously. 

Consensus is not the only way to verify in a trustless system; you can also use a cryptographic proof. A SNARK (Succinct Non-Interactive ARguments of Knowledge) is a cryptographic primitive that allows one party, called the prover, to convince another party, called the verifier, that a given statement is true, without the verifier needing to perform the actual computation.  Cryptographic techniques like Zero-Knowledge Proofs (ZKPs) and verifiable computing enable a party to prove that a computation was performed correctly without revealing the details of the computation itself. These methods provide strong guarantees of correctness while preserving privacy. Despite these benefits, cryptographic methods are extremely computationally expensive, often times adding 10,000x more work on the prover.  While existing SNARKs exist and have been demonstrated to work on smaller neural network models \cite{kang2022scaling}, their proving times are unusable for the large diffusion and language models being deployed today.  Thus, reproducibility of machine learning goes beyond scientific inquiry; rather it also has applications in consensus mechanisms to verify generative machine learning tasks.

\subsection{Why Reproducability in Generative AI is Hard}
As stated previously, in order to have an effective consensus model, and ultimately verify the correctness of machine learning tasks in a distributed, decentralized system, honest nodes must come to agreement and must agree on the same value.  However, in the realm of machine learning and deep learning, the parallel nature of the computations can introduce a slight non-determinism. This is because floating-point operations are not exactly associative, and when executed in parallel, the order in which they are executed can vary. This can cause tiny differences in the results, which may accumulate over time.  Additionally, some GPUs and hardware accelerators have built-in randomness. For example, the NVIDIA Tensor Cores use probabilistic rounding \cite{ootomo2022recovering}, which can make results slightly different even if everything else is kept constant. 

\begin{figure}
\centering
  \includegraphics[width=11.35cm]{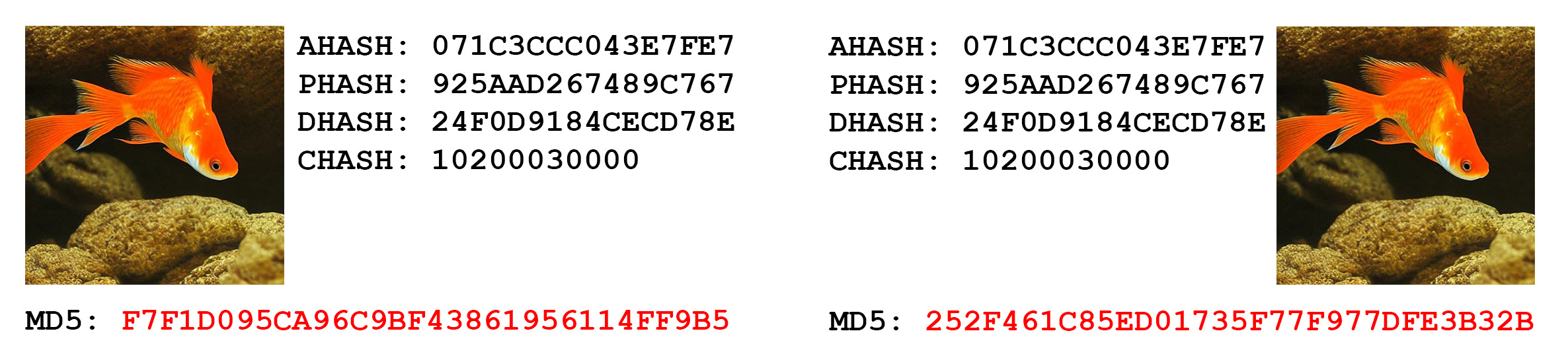}
  \caption{Identical runs of the same seed and prompt to generative AI does not yield the same bit level result, even on the same machine.  Crytographic hashes that are designed for exact matches exhibit the avalanche effect, while perceptual hashes (aHash, pHash, dHash, cHash) are more stable and exhibit locality sensitive hashing.   }
  \label{fig:goldfish}
\end{figure}

Software-wise, machine learning frameworks often introduce non-determinism.  For example, PyTorch, like many other deep learning frameworks, involves operations that can yield different results across multiple executions, even when using identical seeds \cite{reproducability}, see Figure \ref{fig:goldfish}. This non-deterministic behavior can be attributed to factors such as the use of multi-threading, which can lead to race conditions, or specific hardware and software configurations that introduce variability.  While you can limit the sources of non-deterministic behavior or use deterministic algorithms instead of non-deterministic ones where available, this often comes at the cost of performance.  As a concrete example, CUDA convolution operations, which use the cuDNN library, can be a source of non-determinism. This is due to the benchmarking feature of cuDNN, which can select different algorithms for convolution operations based on the size parameters \cite{jorda2019performance}. Disabling this feature can lead to more deterministic but potentially slower performance.

\section{Methodology}
Assuming a set of machine learning nodes are performing inference or training work on generative AI, \textit{how do we know the machine learning node did what it was supposed to do?  Can we reproduce its results?}  Recall in the typical deterministic consensus setting, multiple nodes can re-run the instructions and check the results.  This checking mechanism is typically performed by computing a cryptographic hash of the output or state.  If the instructions were all run correctly, then any validator should have identical state hashes.  Any minor change in the state (like altering a single bit) leads to a substantial change in the output hash, i.e. each of the output bits changes with a 50\% probability. This effect, avalanche effect, can be seen in common hash functions such as the SHA-1 or MD5 hash function. If a single bit is modified, the resulting hash sum becomes entirely different. This makes it extremely difficult to predict the output of the hash function based on a given input, which is a key aspect of its security.

\subsection{Locality-Sensitive Hashing}
In our case, we employ different algorithms that exhibit the Locality-sensitive hashing (LSH) property that probabilistically groups similar input items into the same ``buckets''. Unlike traditional hashing techniques that aim to minimize hash collisions, LSH intentionally maximizes them. This technique can also be viewed as a method for reducing the dimensionality of high-dimensional data, where high-dimensional input items are transformed into lower-dimensional versions while maintaining the relative distances between items.  We utilize the following set of perceptual hashes (a hash string that approximates the visual characteristics of an image), that are commonly used in image hashing,

\noindent\textit{Average Hash (aHash)}: This is a type of perceptual hash that works by resizing the image to a small, fixed resolution, converting it to grayscale, calculating the mean pixel value, and then generating a hash based on whether each pixel is above or below the mean.

\noindent\textit{Perceptual Hash (pHash)}: This is a more complex type of perceptual hash that involves the Discrete Cosine Transform (DCT). The image is resized and converted to grayscale, the DCT is applied, the top-left portion of the DCT matrix (which represents the lowest frequencies) is retained, the mean value is calculated excluding the first element, and a hash is generated based on whether each value is above or below the mean. 

\noindent\textit{Difference Hash (dHash)}: This is another type of perceptual hash that works by comparing the relative gradients of the pixel values. The image is resized and converted to grayscale, each pixel is compared to its neighbor, and a hash is generated based on whether each pixel is greater than or less than its neighbor. 

\noindent\textit{Color Hash (cHash)}: This involves generating a hash based on the colors in an image. It involves resizing the image to a small, fixed resolution and generating a hash based on the quantized color values of each pixel. 

\begin{figure}[ht]
  \begin{center}
  \centerline{
\begin{tabular}{c@{\hspace{1mm}}c@{\hspace{1mm}}c@{\hspace{1mm}}c@{\hspace{1mm}}c@{\hspace{1mm}}c@{\hspace{1mm}}c@{\hspace{1mm}}c}
[469]&
 \parbox[c]{1.5cm}{\includegraphics[width=1.45cm]{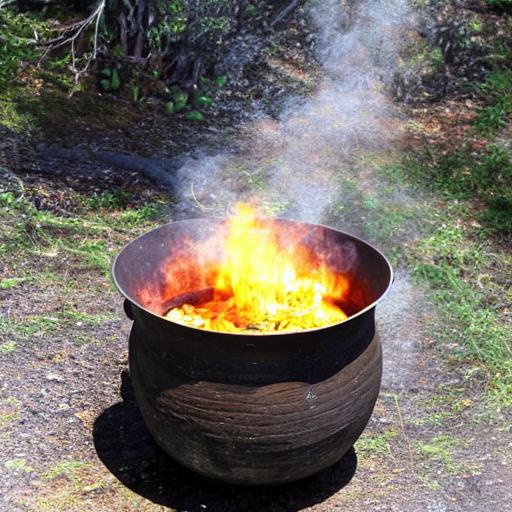}}&
 \parbox[c]{1.5cm}{\includegraphics[width=1.45cm]{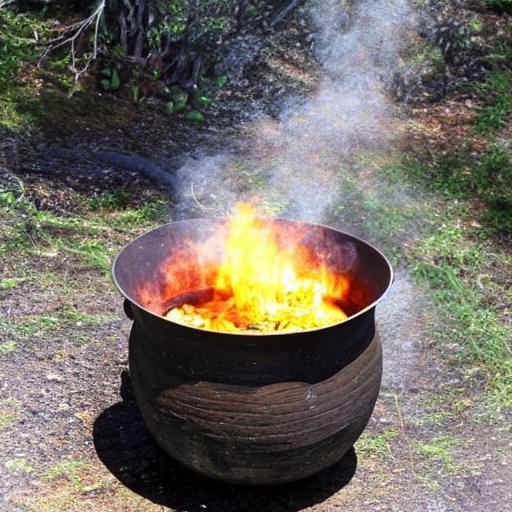}}&
 \parbox[c]{1.5cm}{\includegraphics[width=1.45cm]{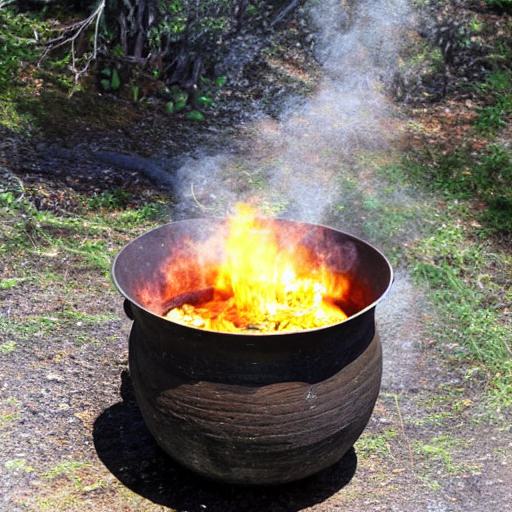}}&
 \parbox[c]{1.5cm}{\includegraphics[width=1.45cm]{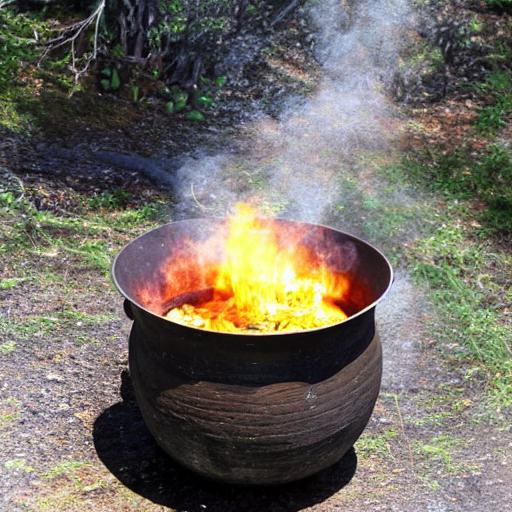}}&
 \parbox[c]{1.5cm}{\includegraphics[width=1.45cm]{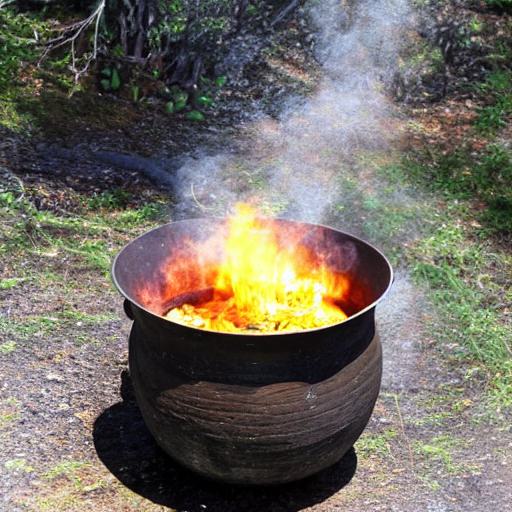}}&
 \parbox[c]{1.5cm}{\includegraphics[width=1.45cm]{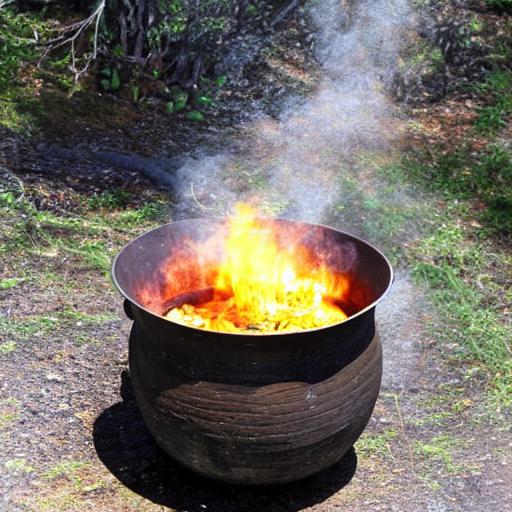}}&
 \parbox[c]{1.5cm}{\includegraphics[width=1.45cm]{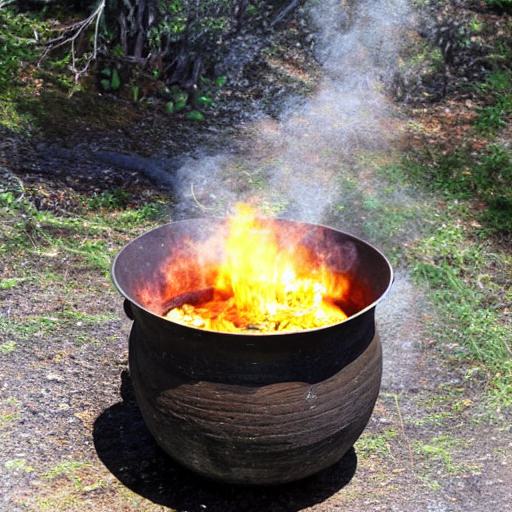}}\vspace{0.1cm}\\ 
 
 [361]&
 \parbox[c]{1.5cm}{\includegraphics[width=1.45cm]{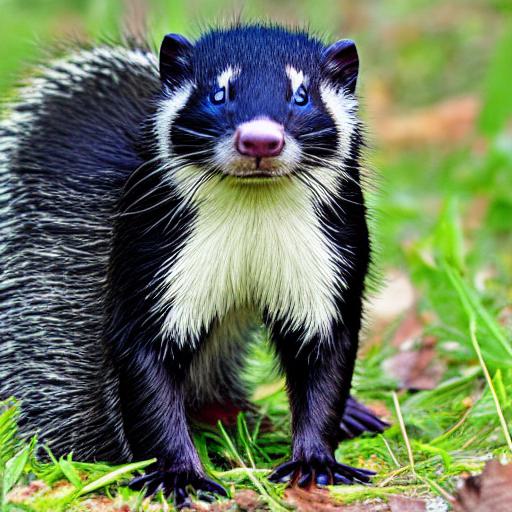}}&
 \parbox[c]{1.5cm}{\includegraphics[width=1.45cm]{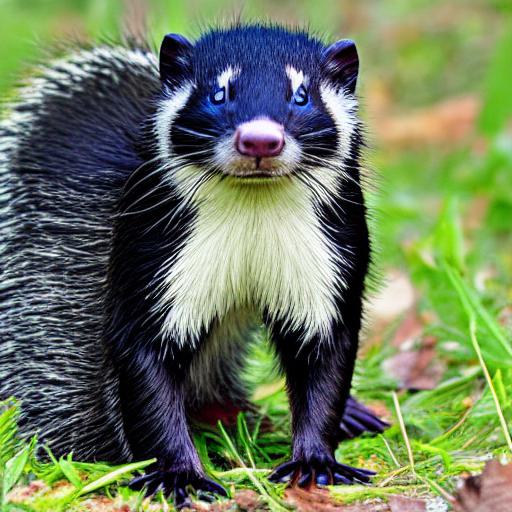}}&
 \parbox[c]{1.5cm}{\includegraphics[width=1.45cm]{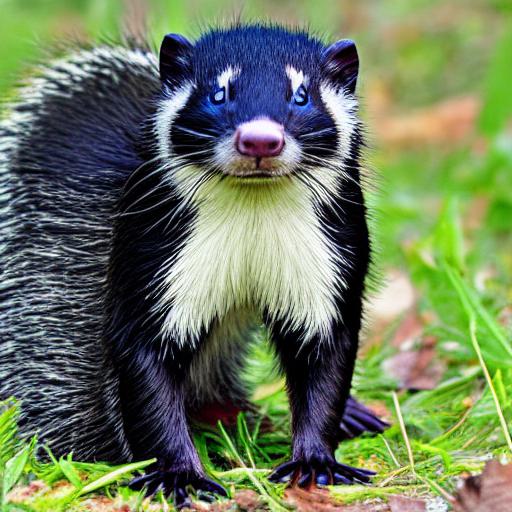}}&
 \parbox[c]{1.5cm}{\includegraphics[width=1.45cm]{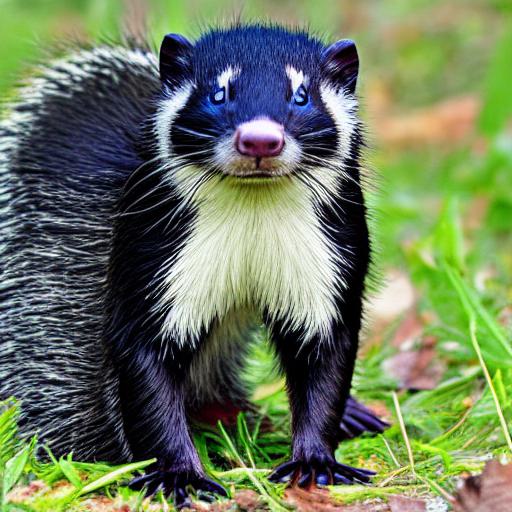}}&
 \parbox[c]{1.5cm}{\includegraphics[width=1.45cm]{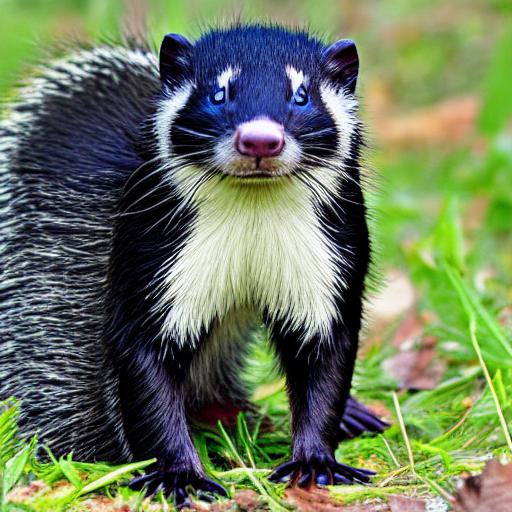}}&
 \parbox[c]{1.5cm}{\includegraphics[width=1.45cm]{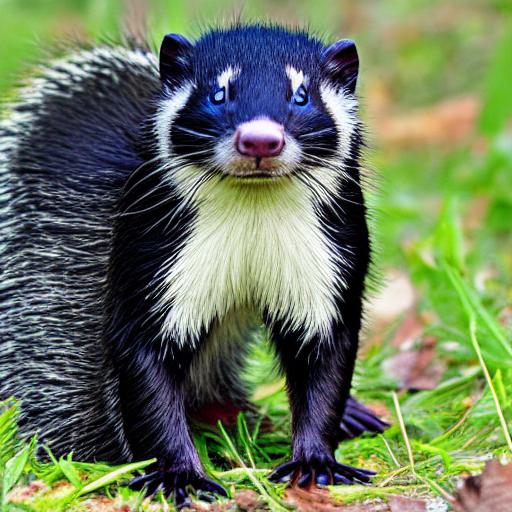}}&
 \parbox[c]{1.5cm}{\includegraphics[width=1.45cm]{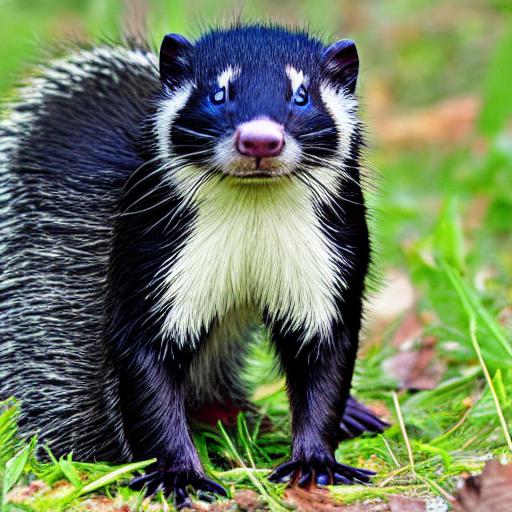}}\vspace{0.1cm}\\ 

 [695]&
 \parbox[c]{1.5cm}{\includegraphics[width=1.45cm]{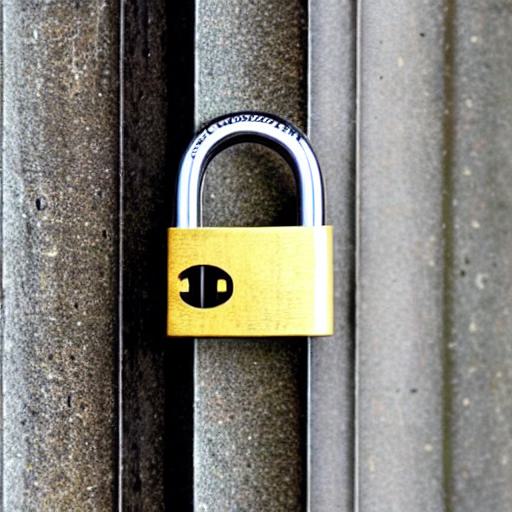}}&
 \parbox[c]{1.5cm}{\includegraphics[width=1.45cm]{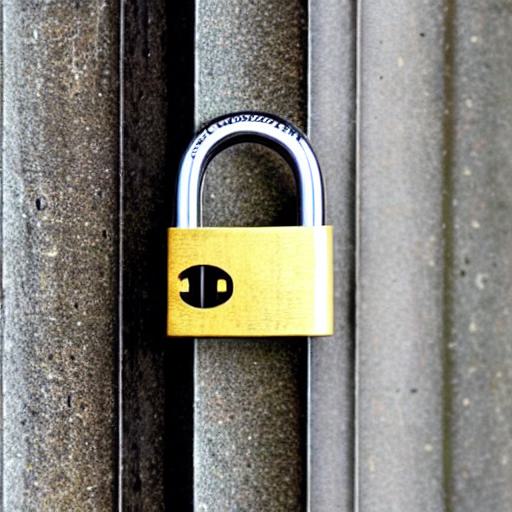}}&
 \parbox[c]{1.5cm}{\includegraphics[width=1.45cm]{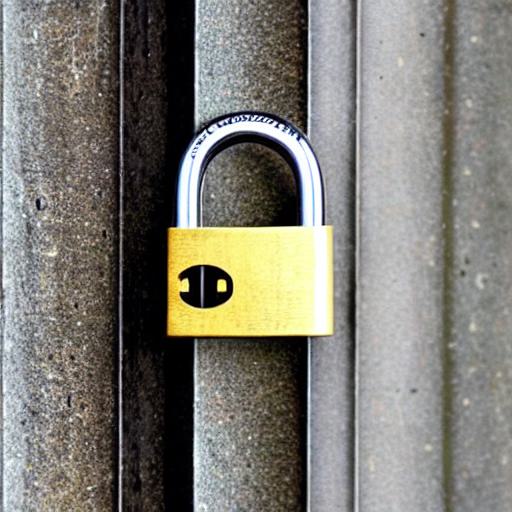}}&
 \parbox[c]{1.5cm}{\includegraphics[width=1.45cm]{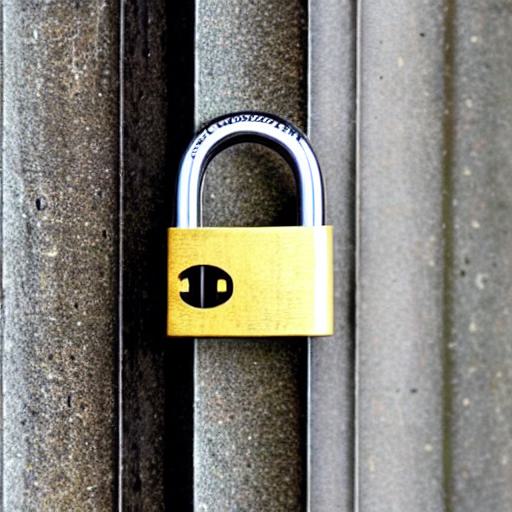}}&
 \parbox[c]{1.5cm}{\includegraphics[width=1.45cm]{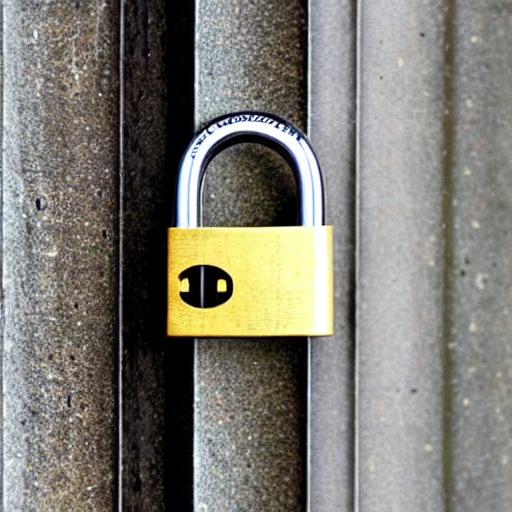}}&
 \parbox[c]{1.5cm}{\includegraphics[width=1.45cm]{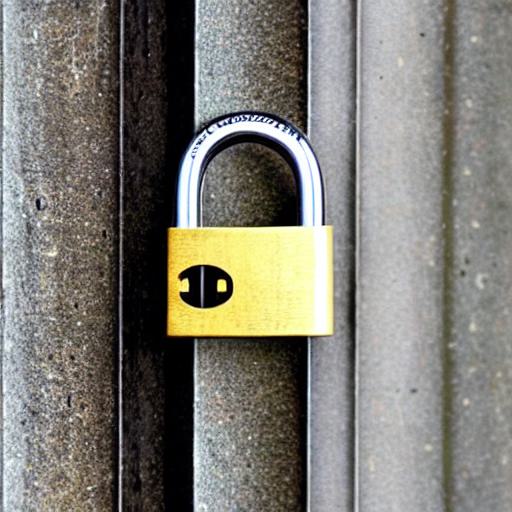}}&
 \parbox[c]{1.5cm}{\includegraphics[width=1.45cm]{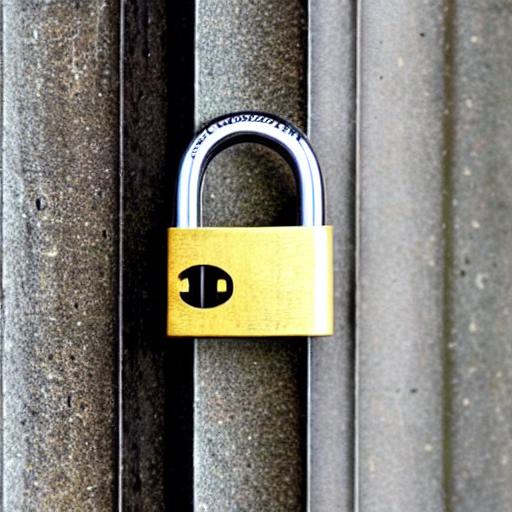}}\vspace{0.1cm}\\ 

[168]&
 \parbox[c]{1.5cm}{\includegraphics[width=1.45cm]{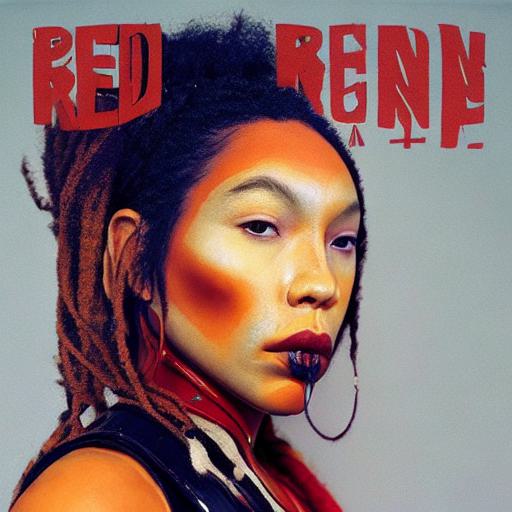}}&
 \parbox[c]{1.5cm}{\includegraphics[width=1.45cm]{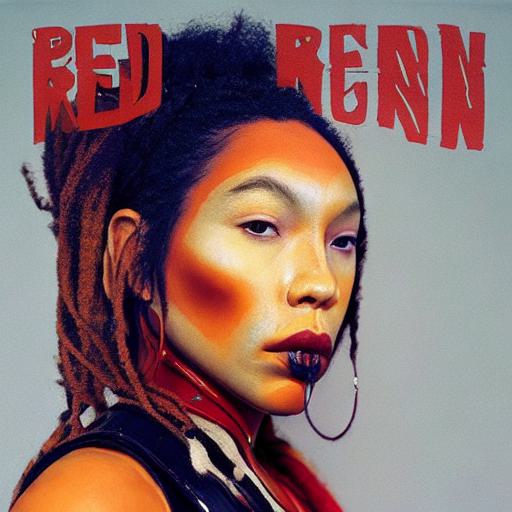}}&
 \parbox[c]{1.5cm}{\includegraphics[width=1.45cm]{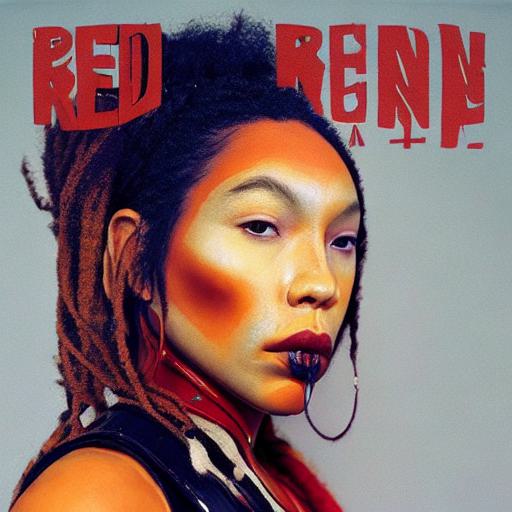}}&
 \parbox[c]{1.5cm}{\includegraphics[width=1.45cm]{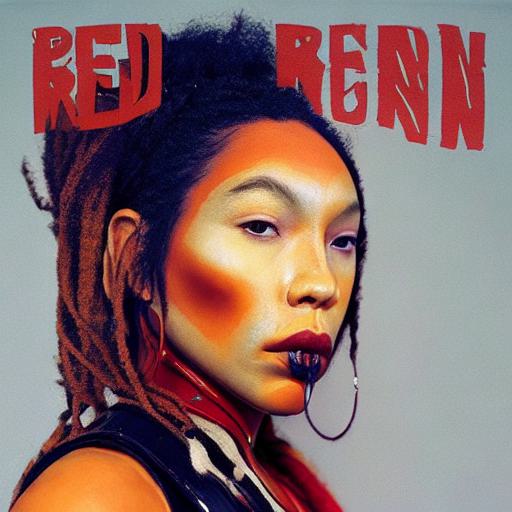}}&
 \parbox[c]{1.5cm}{\includegraphics[width=1.45cm]{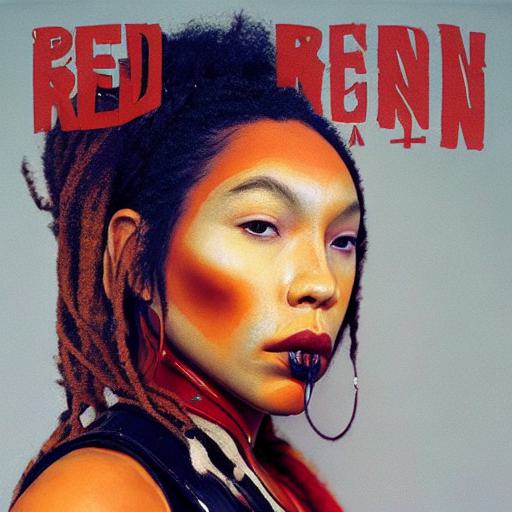}}&
 \parbox[c]{1.5cm}{\includegraphics[width=1.45cm]{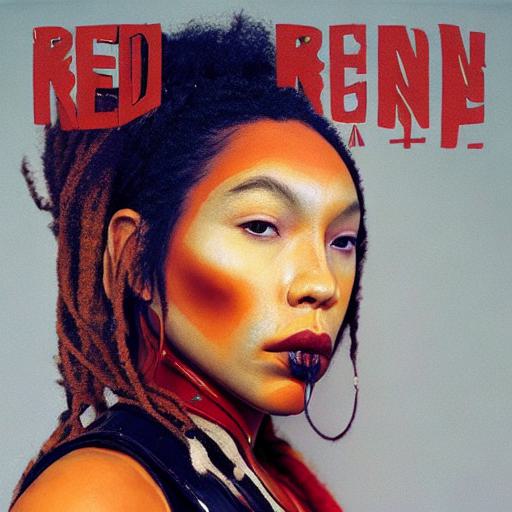}}&
 \parbox[c]{1.5cm}{\includegraphics[width=1.45cm]{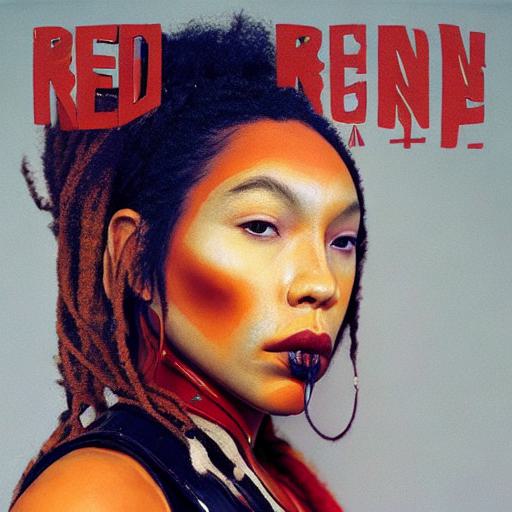}}\vspace{0.1cm}\\ 

 [555]&
 \parbox[c]{1.5cm}{\includegraphics[width=1.45cm]{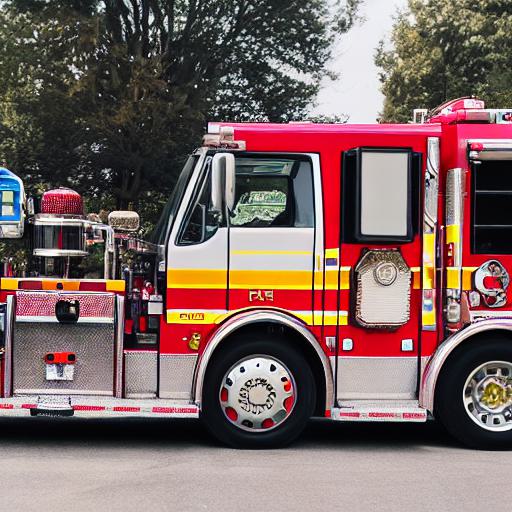}}&
 \parbox[c]{1.5cm}{\includegraphics[width=1.45cm]{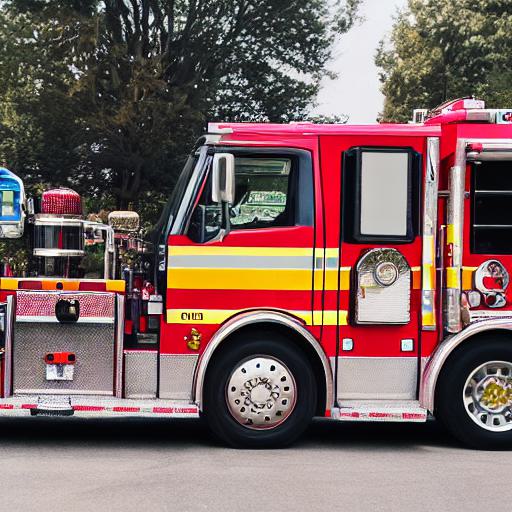}}&
 \parbox[c]{1.5cm}{\includegraphics[width=1.45cm]{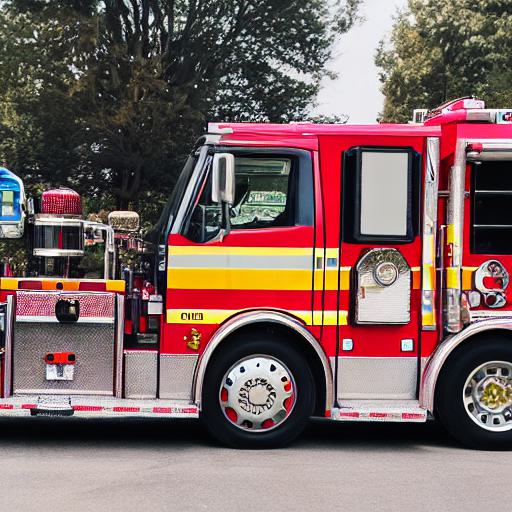}}&
 \parbox[c]{1.5cm}{\includegraphics[width=1.45cm]{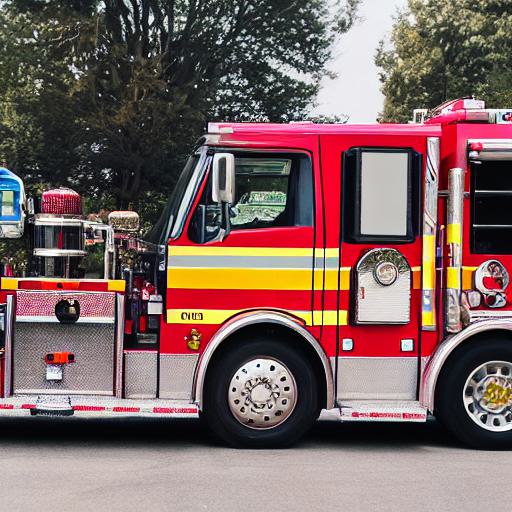}}&
 \parbox[c]{1.5cm}{\includegraphics[width=1.45cm]{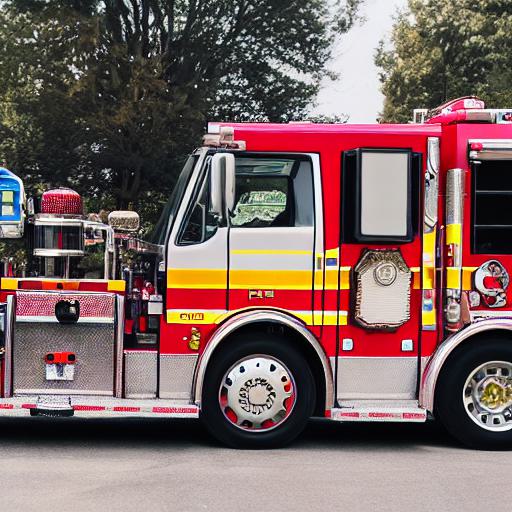}}&
 \parbox[c]{1.5cm}{\includegraphics[width=1.45cm]{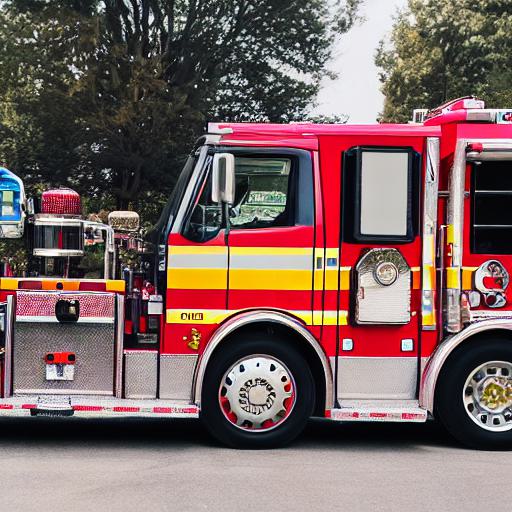}}&
 \parbox[c]{1.5cm}{\includegraphics[width=1.45cm]{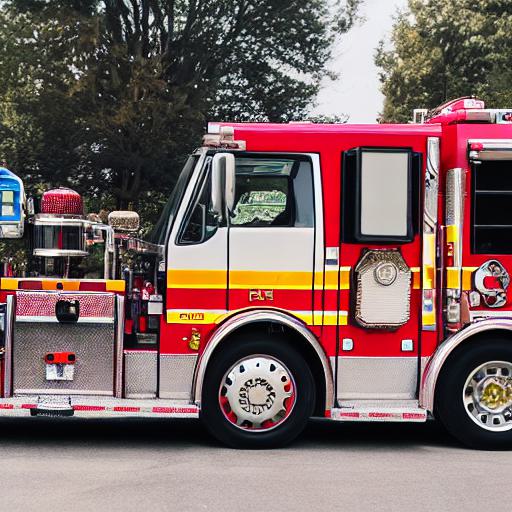}}\vspace{0.1cm}\\ 

  [695]&
 \parbox[c]{1.5cm}{\includegraphics[width=1.45cm]{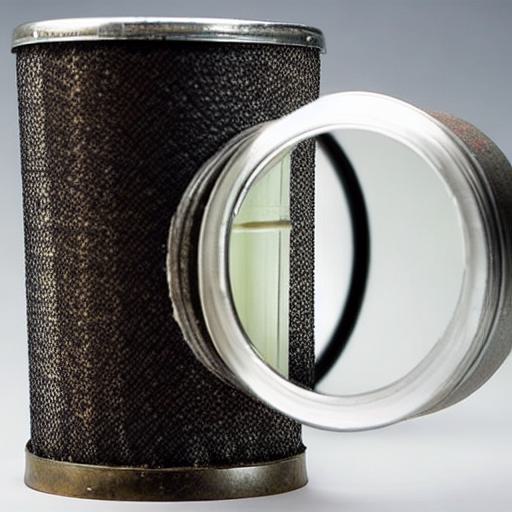}}&
 \parbox[c]{1.5cm}{\includegraphics[width=1.45cm]{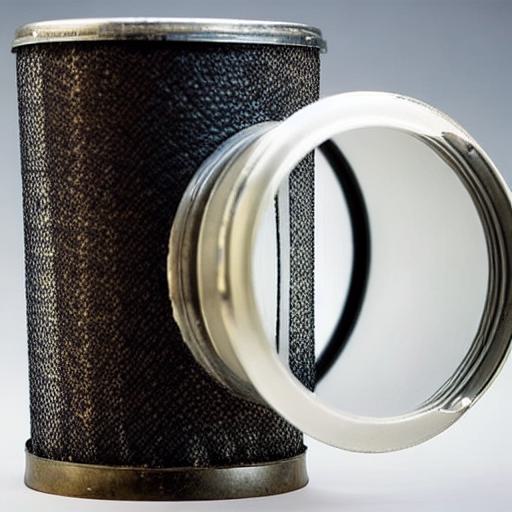}}&
 \parbox[c]{1.5cm}{\includegraphics[width=1.45cm]{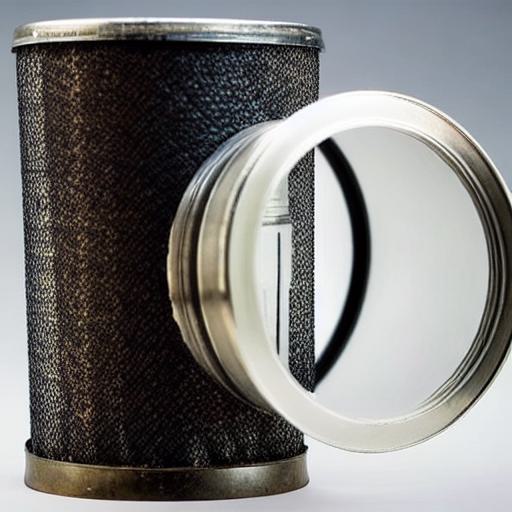}}&
 \parbox[c]{1.5cm}{\includegraphics[width=1.45cm]{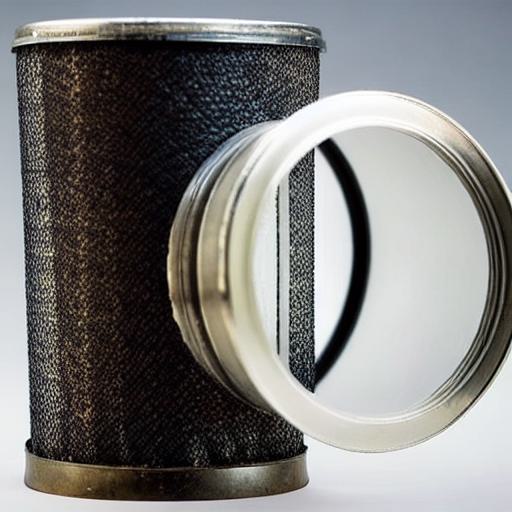}}&
 \parbox[c]{1.5cm}{\includegraphics[width=1.45cm]{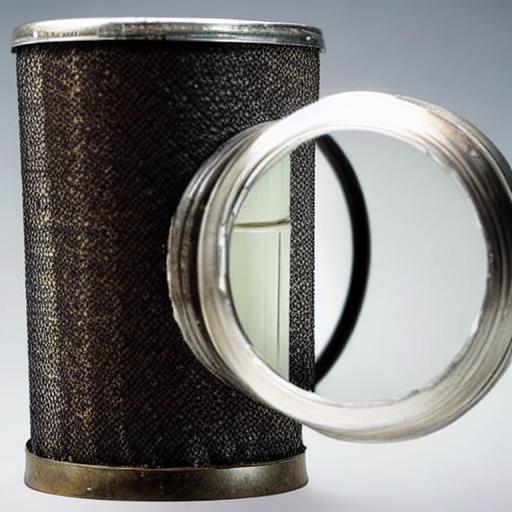}}&
 \parbox[c]{1.5cm}{\includegraphics[width=1.45cm]{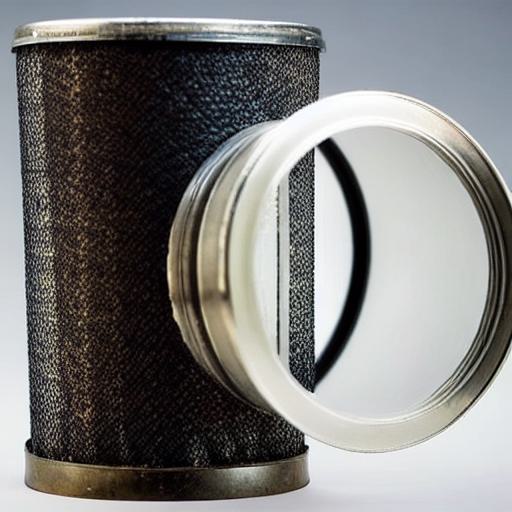}}&
 \parbox[c]{1.5cm}{\includegraphics[width=1.45cm]{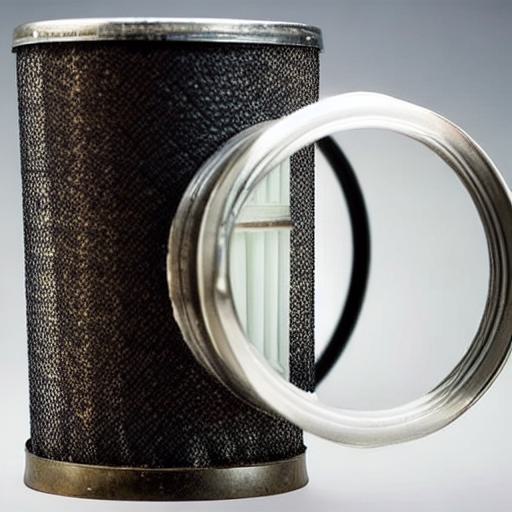}}\vspace{0.1cm}\\ 

 & (a) & (b) & (c) & (d) & (e)& (f) & (g)\\
\end{tabular}}
\caption{Generated images from SD v1.5 on seven different GPUs (mix of 3060ti, 3070ti, 3080ti, and 3090) using the prompt, ``A photo of \{class\}'', where class is from the ImageNet dataset. The first three columns, class id (469) [caldron, cauldron], (361) [skunk, polecat], and (695) [padlock] have identical perceptual hashes.  The last three rows, (168) [redbone], (555) [fire engine, fire truck], (686) [oil filter] have perceptual hashes with the most extreme hamming distances we observed in our generated data (up to 5).  Visual differences can be seen in the writing, the color of the truck, and the circle in the oil filter.}
\label{fig:hashes}
\vspace{-0.5cm}
\end{center}
\end{figure}

These types of perceptual hashes are used in search by image, e.g. Google images searching \cite{testinghash}, or things like identifying songs with the same fingerprint (Shazam) \cite{lsh}.  For our purposes, the perceptual hashes work well, but there are some slight variations in the hash as shown by Figure \ref{fig:hashes}.

\subsection{Generative Image and Language Models}
Our image generation experiments are centered around stable diffusion models and fine-tuned variants.  Stable diffusion supports the ability to generate new images from scratch through the use of a text prompt describing elements to be included or omitted from the output.  The architecture of the model includes a variational autoencoder (VAE), a U-Net, and text encoder. Gaussian noise is iteratively applied to the compressed latent representation during forward diffusion. The U-Net denoises the output from forward diffusion backwards to obtain a latent representation. Finally, the VAE decoder generates the final image by converting the representation back into pixel space.

Popular fine-tuning methods have been employed to personalize the diffusion models, including fine-tuning all of the weights, or low rank and textual fine-tuning methods.   
\subsubsection{Low-Rank Adapation} 
The LoRA, or Low-Rank Adaptation \cite{hu2021lora}, proposes a new method for adapting large-scale pre-trained language models to specific tasks or domains.  Full fine-tuning (retraining all model parameters) becomes less feasible as the size of models continue to expand, i.e. fine-tuning the 175 billion parameters in GPT-3 is prohibitively expensive.

Low-Rank Adaptation (LoRA) freezes the pre-trained model weights and introduces trainable rank decomposition matrices into each layer of the Transformer architecture. This approach significantly reduces the number of trainable parameters for downstream tasks.  LoRA can reduce the number of trainable parameters by orders of magnitude.  In the context of image generation, LoRAs have been used to guide the diffusion process towards particular concepts or visual representations.  

\subsubsection{Textual Inversion}
Textual inversions \cite{gal2022image} are an alternative approach to personalizing text-to-image generation.  Using only a small number of images (typically 3-5 images) of a user-provided concept, such as an object or a style, inversions learn to represent visual concepts through new ``words'' in the embedding space of a frozen text-to-image model. These words can be composed into natural language sentences, guiding personalized creation.  Interestingly, a single word embedding is sufficient for capturing unique and varied concepts. 

\subsection{Large Language Model Generation}
Text generation is crucial for many NLP tasks, including open-ended text generation, summarization, translation, and more. The text generation process, known as decoding, can be customized to improve the quality of the generated output and reduce repetition.  However, in our case, our goal is to produce quality output that can be verified by other nodes.  

The generation configuration that yields deterministic results uses a simple decoding strategy called greedy search, which picks the token with the highest probability as the next token.  For a more globally ``aware'' generation strategy, we can also utilize beam-search decoding, which keeps several hypotheses at each time step and eventually chooses the hypothesis that has the overall highest probability for the entire sequence.

We minimize the stocasticity of the output by turning off stragegies such as multinomial sampling which randomly selects the next token based on the probability distribution over the entire vocabulary, and Beam-Search Multinomial Sampling.

\section{Experiments and Results}
For our experiments, we set up a heterogeneous, decentralized GPU network consisting of (8x) 3080ti GPUs running on 1x PCI-e risers, (4x) 3070ti GPUs running on GPU risers, (4x) 3060ti GPUs, and two other machines each running a single 3090 connected directly to the motherboard via PCI-e 16x slots. There are an additional (4x) A40 Nvidia GPUs connected in an NLink configuration.  In total, the tasks described below were performed on a mix of 22 total GPUs.  Unless specified in the experiment, the GPUs were selected to perform a task at random.  Each machine was running in an Ubuntu 22 environment with CUDA 12 and torch 2.0.

We structure our results as follows:  (1) We first compute the independent likelihoods of determinism in several image and language models.  (2) We then use these likelihoods to compute an algorithm to verify machine learning inference tasks on a decentralized network.  (3) Next, we describe how to verify in the training setting, and lastly (4) share the derived generated dataset for reproducibility and to benefit the community at large.

\subsection{Consensus Likelihoods of Image and Language Generation}
In our first experiment, generate images from different variants of stable diffusion \cite{Rombach_2022_CVPR}, including the base v1.5 model, a fine-tuned version of stable diffusion, LoRA or low rank adapatations \cite{hu2021lora}, and textual inversions \cite{gal2022image}.  We use a template ``A photo of \{\}'', where the class variable inserted is a class from the ILSVRC challenge \cite{deng2009imagenet}.  We generate 7 images with identical prompts and seeds, for a total of approximately 7000 images per image model.  The results of our generation can be seen in Table \ref{tab:results}.

We compute four different types of perceptual hashes and compare the results.  Each class (group of 7 images) is hashed and compared with each other.  The mode hash is selected as the ``correct'' one, and the number of outliers are computed across all thousand categories and reported as the outliers.  We also record the number of outliers when allowing for a tolerance of 1 (Ot$<$=1) or 2 (Ot$<=$2) hash discrepancies (hamming distance of 1 or 2).  Additionally, we report the average hamming distance (aDist) of the outliers and the time to compute a single hash on a 512x512 image.  While the color hash technique yields the best accuracy results, the average hash demonstrates nearly the same performance, yet with a speed up of approximately 3.5x.  The consensus percentages are computed by the number of outliers over the total number of images hashed.  This can be used as the likelihood that an image generated with the same prompt and seed will generate the same perceptual hash.

\begin{center}
\begin{table}[tbh]
    \scriptsize
    \begin{tabular}{ p{3.65cm}  l l l l l l l l }
    \hline
    \textbf{Method}  & \textbf{\# Im.} & \textbf{Consensus} & \textbf{Outli.} & \textbf{Ot<=1} & \textbf{Ot<=2} & \textbf{aDist.}  & \textbf{Time}\\ \hline 
    \multicolumn{1}{c}{Perceptual Hash} \\ \hline
    Stable Diffusion (v1.5) \cite{Rombach_2022_CVPR} & 6985 & 92.9/ 92.9/ 99.7\% & 496 & 496 & 21 & 2.104 & 0.0022s \\ 
    Fine Tuned Diffusion \cite{Rombach_2022_CVPR} & 6997 & 82.5/ 82.5/ 96.7\% & 1224 & 1224 & 233 & 2.615 & 0.0021s\\ 
    Low Rank Adapatation \cite{hu2021lora} & 6928 & 76.9/ 76.9/ 94.1\% & 1598 & 1598 & 410 & 2.871 & 0.0022s\\ 
    Textual Inversion \cite{gal2022image}  & 7035 & 76.4/ 76.4/ 94.8\% & 1657 & 1647 & 365 & 2.931 & 0.0022s\\ 
    \hline
    \multicolumn{1}{c}{Difference Hash} \\ \hline
    Stable Diffusion (v1.5) \cite{Rombach_2022_CVPR} & 6985  & 89.1/ 97.6/ 99.3\% & 764 & 169 & 46 & 1.327 & 0.0020s \\ 
    Fine Tuned Diffusion \cite{Rombach_2022_CVPR} & 6997 & 74.1/ 89.9/ 95.5\% & 1815 & 708 & 312 & 1.772 & 0.0019s\\ 
    Low Rank Adapatation \cite{hu2021lora} & 6928 & 65.5/ 84.9/ 92.4\% & 2386 & 1042 & 525 & 1.940 & 0.0019s\\ 
    Textual Inversion \cite{gal2022image} & 7035 & 71.5/ 88.1/ 94.2\% & 2008 & 840 & 407 & 2.052 & 0.0019s\\ 
    \hline
    \multicolumn{1}{c}{Average Hash} \\ \hline
    Stable Diffusion (v1.5) \cite{Rombach_2022_CVPR} & 6985 & 95.0/ \textbf{99.2}/ \textbf{99.8}\% & 346 & \textbf{55} & \textbf{11} & \textbf{1.210} & 0.0019s \\ 
    Fine Tuned Diffusion \cite{Rombach_2022_CVPR} & 6997 & 86.6/ 95.3/ 98.0\% & 968 & 327 & 140 &  1.638 & 0.0019s\\ 
    Low Rank Adapatation \cite{hu2021lora}  & 6928 & 80.2/ 94.0/ 97.7\% & 1372 & 412 & 157 & 1.545 & 0.0018s\\ 
    Textual Inversion \cite{gal2022image} & 7035 & 82.5/ 94.7/ 98.1\% & 1229 & 370 & 137 & 1.666 & \textbf{0.0018s} \\ 
    \hline
        \multicolumn{1}{c}{Color Hash} \\ \hline
    Stable Diffusion (v1.5) \cite{Rombach_2022_CVPR} & 6985 & \textbf{96.8}/ 98.5/ 99.5\% & \textbf{227} & 105 & 37 & 1.700 & 0.0074s \\ 
    Fine Tuned Diffusion \cite{Rombach_2022_CVPR} & 6997 & 92.1/ 96.2/ 98.8\% & 555 & 265 & 86 & 1.754 & 0.0073s\\ 
    Low Rank Adapatation \cite{hu2021lora} & 6928 & 90.6/ 95.6/ 98.7\% & 652 & 303 & 93 & 1.719 & 0.0082s\\ 
    Textual Inversion \cite{gal2022image} & 7035 & 89.2/ 94.6/ 98.2\% & 759 & 379 & 124 & 1.760 & 0.0070s\\ 
    \hline
    \end{tabular}
    	\caption{Generated images from different variants of stable diffusion \cite{Rombach_2022_CVPR}, including the base v1.5 model, a fine-tuned version of stable diffusion, LoRA or low rank adapatations \cite{hu2021lora}, and textual inversions \cite{gal2022image}.  Approximately 7 images with identical prompts and seeds were generated per 1000 classes in the ImageNet class categories.  The number of outliers to the majority hash in the class is shown with tolerances of 0, 1 (Ot$<$=1), and 2 (Ot$<$=2).  The average distance of all outliers is shown as aDist, and the time per image hash is presented in the last column. }
	\label{tab:results}
	\end{table}
\end{center}

For language generation, we sample several open source LLMs and run the same prompt over four different machines and GPUs.  The five LLMs tested were the Wizard Vicuna 13B \cite{xu2023wizardlm} in 4-bit quantizaiton mode, the Vicuna 7B \cite{vicuna2023} in 8-bit quantization, Red Pajama 7B \cite{together2023redpajama} in 8-bit quantization, the Red Pajama 3B \cite{together2023redpajama}, and GPT-J 6B \cite{mesh-transformer-jax} in 8-bit quantization.  We utilized the ``instruct'' version of the LLM and provided a prompt, \textit{``\#\#\# Human: Please write a description about \{name\} \#\#\# Assistant:''},  where the name comes from the 1000 ImageNet classes.  A total of 4000 sentence generations were created across different decoding techniques and beam searches as shown in Table \ref{tab:llms}.  Even with the GPU non-determinism, we saw no stocasticity in the LLM outputs when greedy or n-beam methods were used to decode.  When explicitly specifying multinomial sampling where the model selects the next token based on the probability distribution over the entire vocabulary, we do observe the expected non-deterministic behavior.  As a final note, we believe that although we did not observe any mismatched tokens between GPUs, we believe that there is a non-zero chance of a token mismatch in the greedy and n-beam case.  The nature of the floating point operations and error drift, with a special edge case probability that is near the border of two words, could possibly flip the generated next word; however, this would be an extremely rare edge case.
\begin{center}
\begin{table}[tbh]
    \scriptsize
    \begin{tabular}{ p{3.05cm}  l l l l l l l }
    \hline
    \textbf{Method}  & \textbf{\# Gen.} & \textbf{Greedy(30)} & \textbf{Greedy(60)} & \textbf{Beam N=5} & \textbf{Beam N=10} & \textbf{Multin.} \\ \hline 
    Wizard Vicuna 13B \cite{xu2023wizardlm} & 4000 & Yes & Yes & Yes & Yes & No \\ 
    Vicuna 7B \cite{vicuna2023} & 4000 & Yes & Yes & Yes & Yes & No \\ 
    Red Pajama 7B \cite{together2023redpajama} & 4000 & Yes & Yes & Yes & Yes & No \\ 
    Red Pajama 3B \cite{together2023redpajama} & 4000 & Yes & Yes & Yes & Yes & No \\ 
    GPT-J 6B \cite{mesh-transformer-jax} & 4000 & Yes & Yes & Yes & Yes & No \\ 
    \hline
 
    \end{tabular}
    	\caption{Generated 4,000 descriptions from different LLMs of different parameter sizes.  Different decoing strategies include greedy methods of 30 max tokens and 60 max tokens, beam search of size 5 and 10, as well as multinomial decoding.  All greedy and beam search strategies were empirically deterministic across different machines and GPUs.  The multinomial case was non-deterministic as expected.  A ``Yes'' indicates observed deterministic behavior. }
	\label{tab:llms}
	\end{table}
 \vspace{-0.5cm}
\end{center}

\subsection{Tolerance in Reproducability}
\begin{figure}[tbh]
\centering
  \subfigure[Simple Majority]{\includegraphics[width=6.05cm]{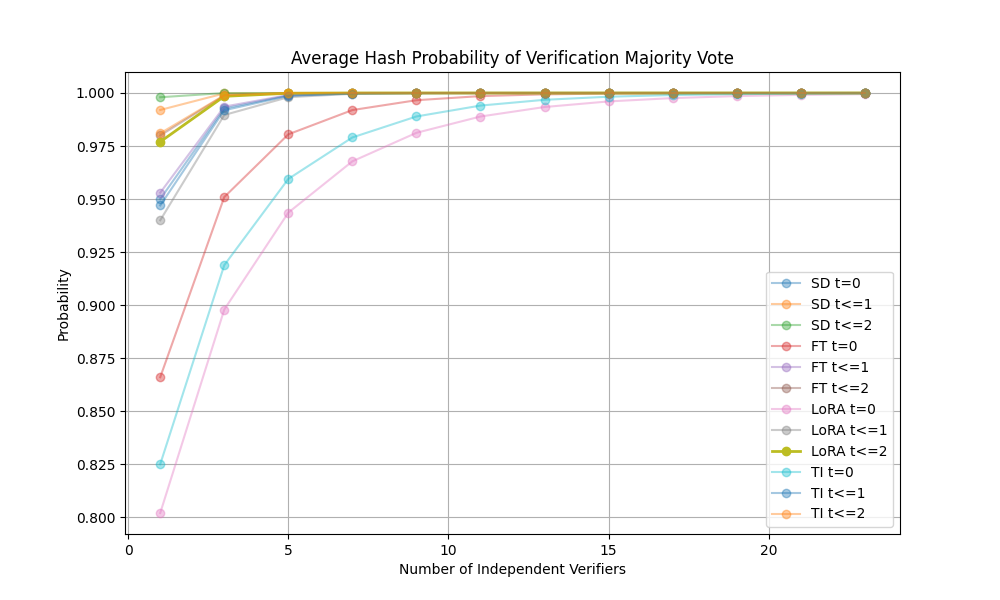}}
    \subfigure[2/3 Super Majority]{\includegraphics[width=6.05cm]{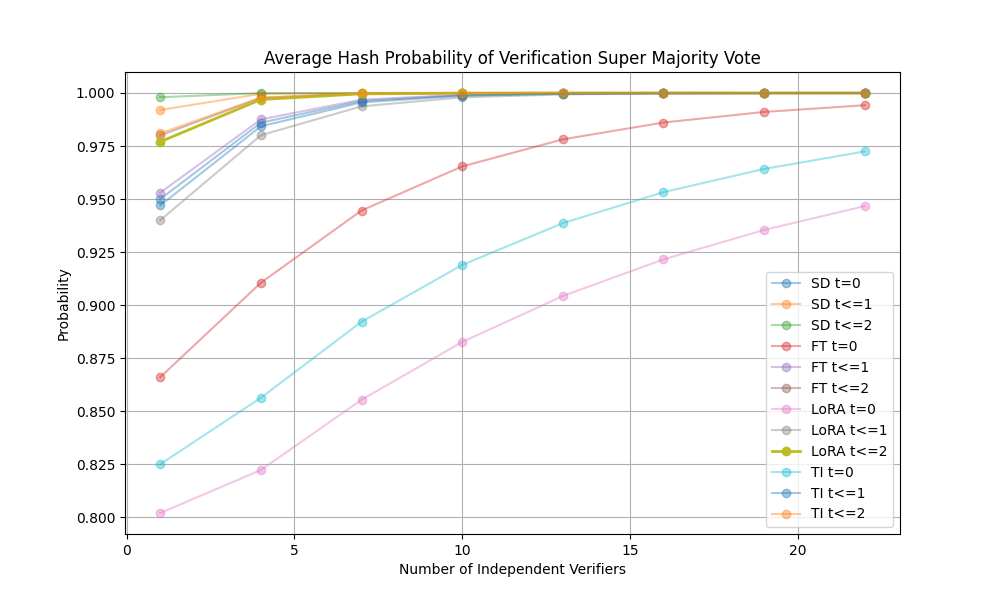}}
  \caption{Graphs of probabilities that independent verifiers can spot incorrect or fraudlent behavior given different likelihoods of deterministic generation.  In the simple majority case, we assume a majority of nodes are honest, and in the super majority case, we have a stricter assumption that over 2/3s are honest.}
  \label{fig:maj}
\end{figure}

Given the independent likelihoods of determinism computed from the previous experiment, we can now derive an machine learning algorithm for a decentralized network.  We provide posterior probabilities of detecting incorrect or fraudulent behavior in the case that we assume the majority of nodes in the network are honest (Figure \ref{fig:maj}(a)), or the case where we require a super-majority (Figure \ref{fig:maj}(b)).  We use, $P(X=k) = \binom{n}{k} p^k (1-p)^{n-k}$
, where the binomial distribution gives the probability of getting exactly $k$ successes (defined as generating the correct perceptual hash) over $n$ independent verifiers.  The $P(X=k)$ is the probability of getting exactly k verifications, where $p$ is the probability of generating a correct output (derived from above).  

\noindent\textbf{Verification of Correctness - Type I Error} - For majority vote and tolerance of 2, we can achieve 99.843\%, 99.988\%, and 99.999\% verification with 3, 5, and 7 independent verifiers.  For super majority (greater than 2/3s) and tolerance of 2, we can achieve 99.692\%, 99.960\%, and 99.999\% verification accuracy with 4, 7, and 10 independent verifiers; these probabilities demonstrate strong verification with a minimal redudant work.

Another type of error is the accidental or malicious generation of a perceptual hash without performing the task, or guessing a hash based upon the given prompt.  To assess the risk of this error, we generate over 1.3M images using stable diffusion.  Our simulation mimics a scenario where a malicious actor sees the prompt, and tries to guess at a perceptual hash.  Thus, these guess images are generated using identical prompts as the one provided (a photo of \{\}), but with a different seed.  

\noindent\textbf{Verification of Correctness - Type II Error} - For each category, we generated 1300 images (a total of 1.3M), and perform an all-to-all average hash comparison and count collisions.  The total number of hash comparisons here is approximately 0.85 billion.  The probability of an intra-class perceptual hash collision is less than 0.0267\%, when allowing for a hamming distance tolerance of $<$=2.  This indicates that there is nearly zero percent chance that an adversary would be able to guess the perceptual hash - even with information about the prompt given.

\subsection{Verification of Generative Training}
We now turn our attention to the case of generative AI training.  In particular, we focus on the textual inversion fine-tuning case; this is a likely scenario in a distributed, heterogeneous GPU network.  Textual Inversion is a technique for capturing novel concepts from a small number of example images. It learns new ``words'' in the text encoder's embedding space, which are used within text prompts for personalized image generation.  Importantly, the weights of the U-Net and VAE are frozen, thus, the only parameters that are able to change and reduce the loss are the 768-dimensional word embedding.

\begin{figure}
\centering
  \includegraphics[width=12.5cm]{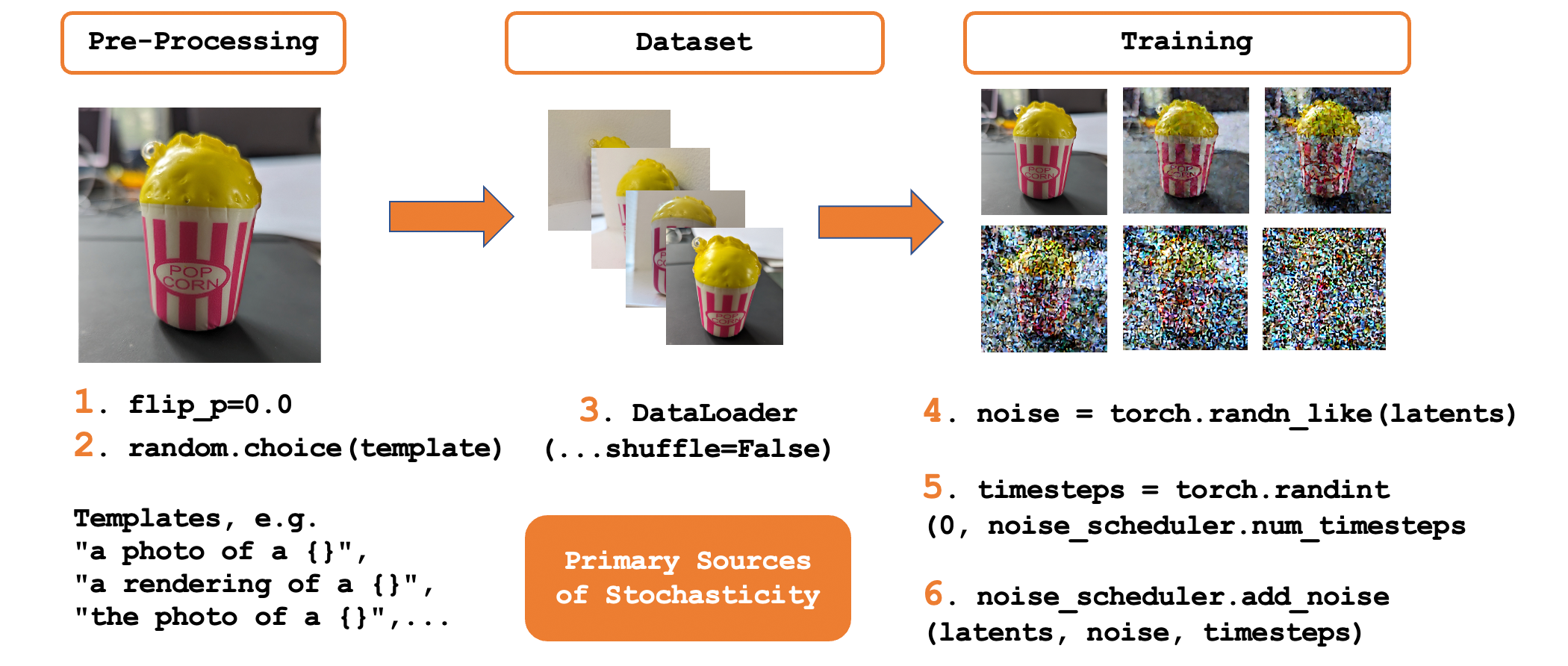}
  \caption{Six major sources of stocasticity in the fine-tuning process of textual inversion.  The randomness is minimized at each of these six points in order to gain verifiability in the training process.  Even with control over these parameters via seeds or parameter settings, the process still remains non-deterministic. }
  \label{fig:six}
\end{figure}

Similar to the inference case, we would need the ability to perform a fraud proof over training epochs.  Thus, we need to identify the major sources of controllable stocasticity and minimize them to allow for replication.  Figure \ref{fig:six}, identifies and minimizes the six controllable sources of randomness including, horizontal flips of the data, random choice of the template, training data shuffling, gaussian noise for the latents, random choice of timestep, and the noise scheduler.

\begin{figure}
\centering
  \subfigure[$<$popcorn-toy$>$]{\includegraphics[width=3.95cm]{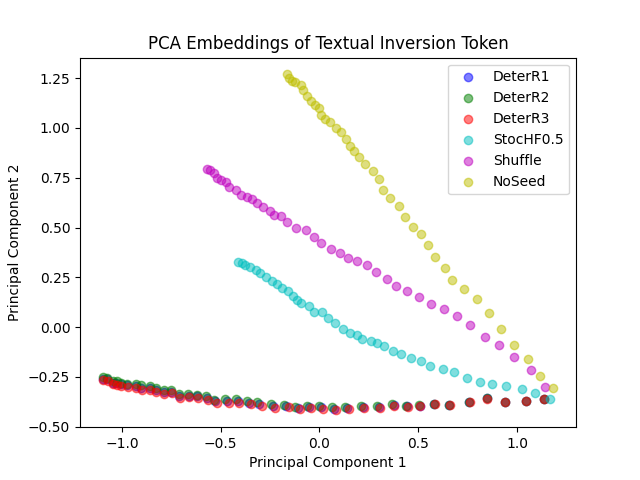}}
    \subfigure[$<$kitchenrobot$>$]{\includegraphics[width=3.95cm]{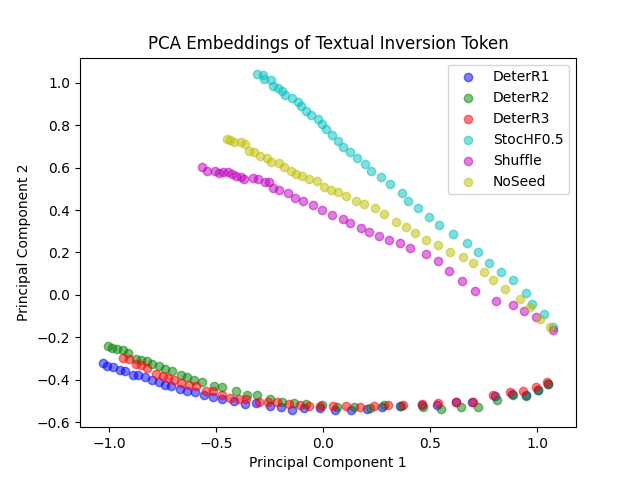}}
    \subfigure[$<$ugly-sonic$>$]{\includegraphics[width=3.95cm]{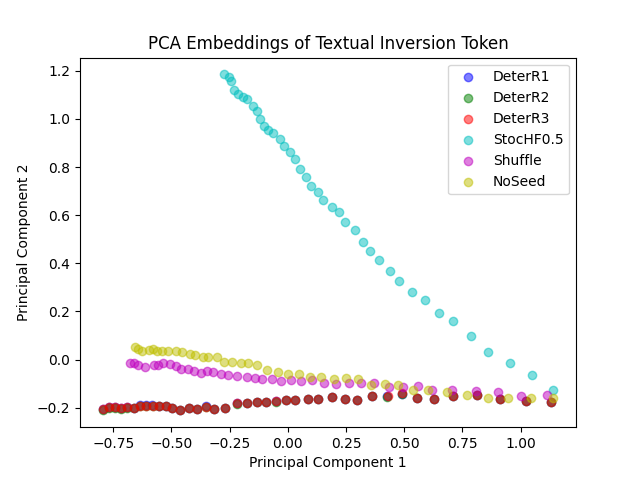}}
    \subfigure[$<$clothes$>$]{\includegraphics[width=3.95cm]{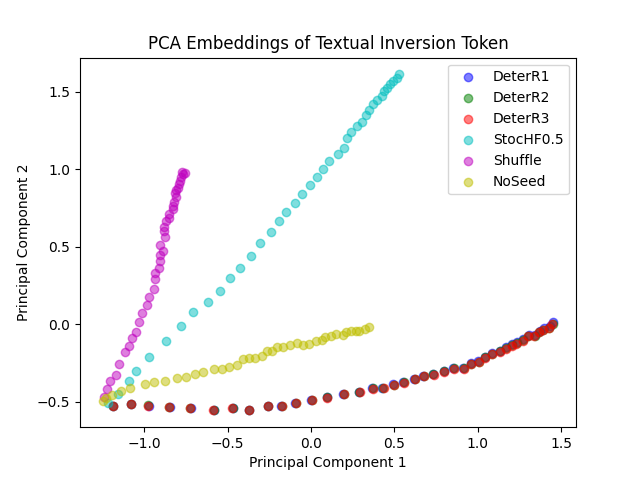}}
    \subfigure[$<$kirby$>$]{\includegraphics[width=3.95cm]{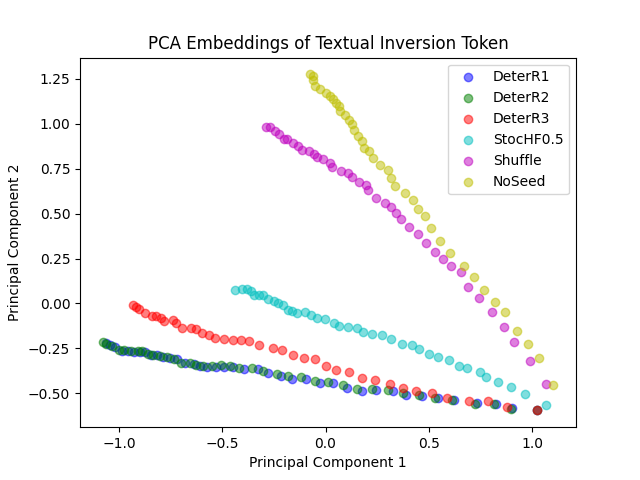}}
    \subfigure[$<$car-toy$>$]{\includegraphics[width=3.95cm]{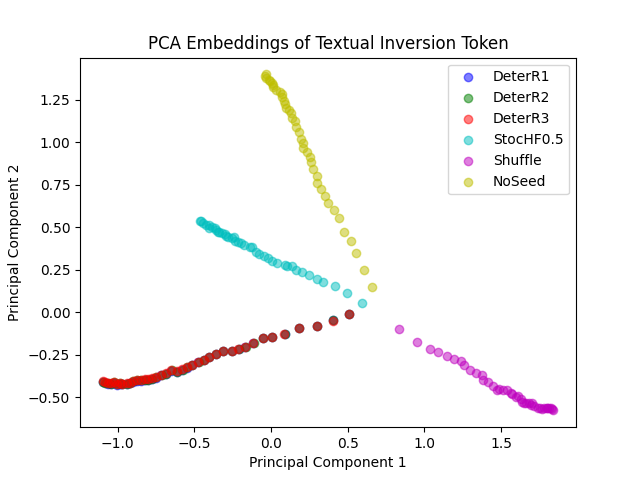}}
  \caption{Textual inversion training results from randomly selected objects in the sd concepts library (https://huggingface.co/sd-concepts-library).  The graphs show six runs of 2000 iterations fine-tuning a new token based upon a small set of images (3-10).  Three runs minimize the amount of stocasticity within the training process, ``DeterR1,2,3''.  ``StocHF0.5'' includes a random horizontal flip of the image, ``Shuffle'' allows the dataset to be shuffled, and ``Noseed'' does not provide a random seed for the noise.  The deterministic runs mostly overlap with some drift over the plot of the 768-dimensional vector projected on the 2 principal components of the training data. }
  \label{fig:pca}
\end{figure}

We remove these sources of stocasticity and train several dozens of textual inversions randomly sampled from the huggingface concepts library \footnote{https://huggingface.co/sd-concepts-library}.  For each concept, we present 5-10 exemplar images, and train the model with the following hyperparameters: learning rate is 5e-04, maximum train steps is 2000, batch size is 4, gradient accumulation is 1, and checkpoints are generated every 50 steps.  The model is trained in mixed 16-bit precision.  Each textual inversion train is performed six times.  The first three times are deterministic runs - we minimize all controllable sources of stochasticity.  The next three runs are ablations over the possible sources of randomness.  For run 4, we allow horizontal flips in the training process, for run 5 we allow the input dataset to be shuffled, and for run 6, we do not set the seed for training and noise.  We take the six runs and perform PCA over the 768-dimensional space and project the checkpoint embeddings onto principal components 1 and 2.  The plots of the training checkpoints of all six runs can be seen in Figure \ref{fig:pca}.

\begin{SCfigure}
  \centering
  \caption{Training in a textual inversion with resyncing of checkpoints every 300 iterations on the $<$kitchenrobot$>$ inversion. (1) Three training instances (Sync Deter1,2,3) are introduced and synchronized to DeterR1 during training.  (2) At the start, there is some amount of variation within the training but within an error bounds of all three Deterministic Runs.  (3) By the end of training 2000 iterations, the Synchronized trainings and Deterministic R1, are nearly identical. The graph shows 4 overlapping vector projections from the DeterR1 and Sync Deter1,2,and 3.}
  \includegraphics[width=0.5\textwidth]{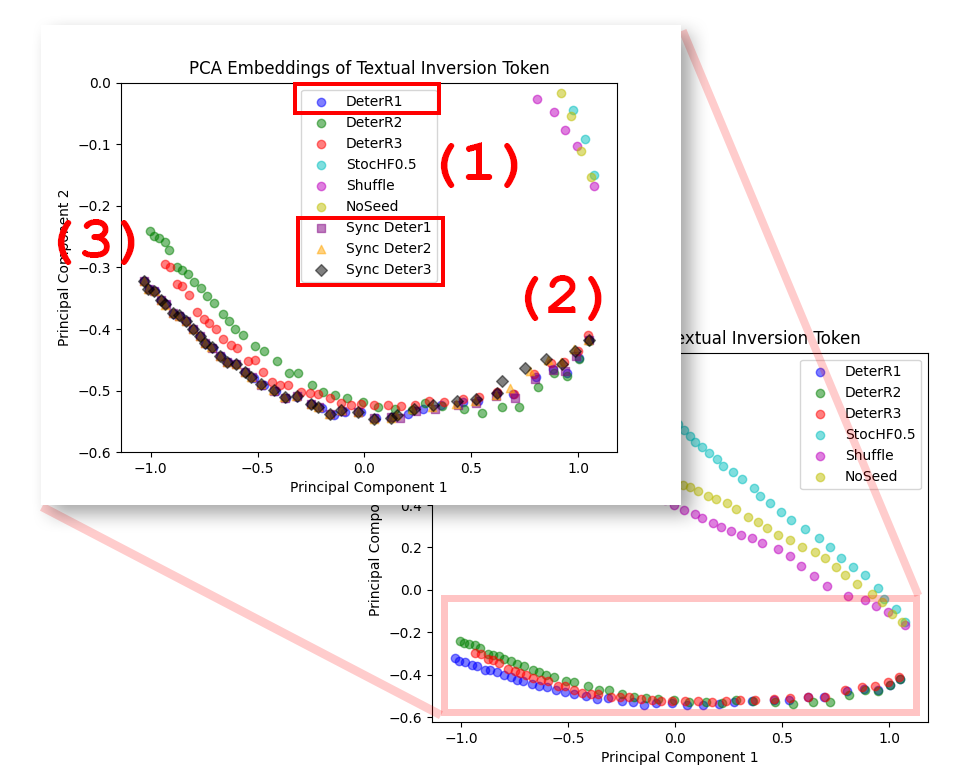}
  \label{fig:gossip}
\end{SCfigure}

We observe that by minimizing the stocasticity leads to nearly deterministic training.  There are cases in Figure \ref{fig:pca}(b) and  Figure \ref{fig:pca}(e) where the random error begins to propagate and cause drift in the final textual embeddings.  To control for this, we demonstrate a gossip training procedure in Figure \ref{fig:gossip} where the checkpoints between deterministic nodes can be synchronized every several checkpoints (here we do every 6 checkpoints, or 300 epochs).  Given the gossip training mechanism, the end embedding weights of the verifiers follow the trajectory of the trainer and provide much tighter error bounds.

\section{Conclusion}
In conclusion, we demonstrate that we are able to reproduce and verify the correctness of machine learning tasks in a distributed, decentralized network.  We tackle a particularly challenging task in ML of verifying both the inference and training of generative AI.  We present the likelihoods of an honest node producing an output that can be verified by other independent nodes in the network and in the process generated millions of data samples and billions of hash comparisons.

We provided empirical evidence of different parameterizations of tolerance and error bounds for verification as the groundwork for a fraud proof in a blockchain network.  Our results show that with minimal overhead and extremely high precision, we can verify generative AI work in both image and language generation, and a malicious actor has close to zero percent chance of exploiting the algorithm. The study also identified and minimized the major sources of stochasticity in generative AI training and presented gossip and synchronization training techniques for verifiability. 

 In summary, we provide a robust and practical foundation for AI verification, reproducibility, and consensus, significantly reducing the need for trust in a decentralized network.





\bibliographystyle{unsrt}
\footnotesize
\bibliography{references}

\end{document}